\documentstyle[12pt]{article}
\def\bc{\begin{center}}
\def\ec{\end{center}}
\def\be{\begin{equation}}
\def\ee{\end{equation}}
\def\bea{\begin{eqnarray}}
\def\eea{\end{eqnarray}}

\def\simge{\ \lower-
1.2pt\vbox{\hbox{\rlap{$>$}\lower5pt
\vbox{\hbox{$\sim$}}}}\ }

\begin{document}
\pagestyle{empty} 
\vspace{-0.6in}
\begin{flushright}
\end{flushright}
\vskip 2.0in
\centerline{\large {\bf{LATTICE GAUGE FIXING, GRIBOV COPIES}}}
\centerline{\large {\bf {AND}}}
\centerline{\large {\bf {BRST SYMMETRY}}}
\vskip 1.0cm
\centerline{M. Testa$^{1,2}$}
\centerline{\small $^1$  Theory Division, 
CERN, 1211 Geneva 23,
Switzerland$^{\star}$.}
\centerline{\small $^2$ Dipartimento di 
Fisica, Universit\`a di Roma ``La
Sapienza"}
\centerline{\small Sezione INFN di Roma}
\centerline{\small P.le A. Moro 2, 00185 
Roma, Italy$^{\star}$$^{\star}$.}
\vskip 1.0in
\abstract{We show that a modification of the BRST lattice
quantization allows to circumvent an old paradox, formulated
by Neuberger, related to lattice Gribov copies and
non-perturbative BRST invariance. In the continuum limit the usual
BRST formulation is recovered.}
\vskip 1.0in
\begin{flushleft} 
\end{flushleft}
\vfill
\noindent \underline{\hspace{2in}}\\
$^{\star}$ Address until August 31st, 1998.

\noindent $^{\star}$$^{\star}$ Permanent address.
\eject
\pagestyle{empty}\clearpage
\setcounter{page}{1}
\pagestyle{plain}
\newpage 
\pagestyle{plain} \setcounter{page}{1}
\section{Introduction}

Besides being a landmark in our understanding of non-perturbative
hadron dynamics, the lattice regularization of gauge theories\cite{wil}
is an important laboratory where formal field theoretical ideas
can be cross-checked for their validity or possible flaws may be identified.

One of these ideas is the gauge fixing. In fact we are in the fortunate
position that lattice regularization does not require any gauge-fixing,
so that the steps towards a gauge-fixed version of the theory are completely
non formal. In particular a non formal meaning could be attached to the
Faddeev-Popov procedure\cite{faddeev}, so essential for many applications.
The Faddeev-Popov procedure has afterwards been replaced by the more formal
apparatus of the so-called BRST symmetry\cite{brs}.
Despite these important developments, some problems are still open.
It is in fact known\cite{grib,sing} that compact gauge fixing
is affected by the presence of Gribov copies and one is therefore faced to
the problem of dealing, in the functional integral, with several gauge
equivalent copies of the same configuration. An appealing solution\cite{sol},
naturally suggested by the BRST formulation, is that the various Gribov
copies, weighted by the Faddeev-Popov determinant, contribute to the
functional integral with alternating signs. This opens the way to a
cancellation in which, finally, only the contribution from a single copy
should survive.
It was shown by Neuberger\cite{neub1}, on very general grounds,
that a cancellation takes place indeed, with the disastrous result of leaving 
physical observables in the embarrassing indeterminate form
${0 \over 0}$. 
The situation becomes even more confused when, as it happens
for chiral gauge theories, a gauge invariant discretization is missing.
One possible scheme for their quantization, known as the Rome
approach\cite{rome}, requires gauge-fixing and BRST
quantization as fundamental ingredients.
While the validity of this approach can be checked within
perturbation theory, most of the objections raised against it\cite{neub2}
concern the lack of control of the gauge fixing procedure at the
non-perturbative level, although Neuberger's argument
is not directly relevant in this case.

In this paper we will show how the paradox presented in ref.\cite{neub1}
can be avoided, through a slight modification of the BRST scheme,
in the case of theories admitting a gauge invariant discretization.

In section \ref{first} we will review the Neuberger's argument.
In section \ref{second} we will clarify its Gribov-like nature through a very
simplified one dimensional integral. In section \ref{third} we will discuss
the modification required in order to obtain a consistent BRST quantization.

\section{The paradox} \label{first}

Following ref.\cite{neub1} we consider a discretized gauge-invariant
theory in a finite volume, so that the functional integral reduces to
a finite-dimensional one. 
The expectation value of any gauge invariant operator ${\cal O}(U)$ is
given by:
\bea
& & \langle {\cal O} \rangle ={1 \over {\cal Z}} \int {\cal D}U e^{-S(U)}
{\cal O}(U) \label{uno}\\
& & {\cal Z}\equiv \int {\cal D}U e^{-S(U)} \nonumber
\eea
where ${\cal D}U$ denotes the
group-invariant integration measure over links and $S(U)$ is the gauge
invariant euclidean action.

The process of gauge-fixing requires, first of all, the choice of
a gauge condition, i.e. a gauge non invariant function, $f(U,x)$,
and the introduction of auxiliary degrees of freedom: the grassmannian
ghost and anti-ghost variables, $c$ and $\bar c$, and the Lagrange
multipliers $\lambda$, to be integrated over the whole real axis. 

After gauge-fixing, the expectation value in eq.(\ref{uno}),
is replaced by:
\bea
& & \langle {\cal O} \rangle =
{1 \over {\cal Z'}} \int {\cal D}\mu \ e^{-S(U)} e^{-{\alpha \over 2}\int
\lambda^2} \ \exp(\delta \int f \bar c) \
{\cal O}(U) \label{due}\\
& & {\cal Z'}\equiv \int {\cal D}\mu \ e^{-S(U)} e^{-{\alpha \over 2}\int
\lambda^2} \ \exp(\delta \int f \bar c) \nonumber
\eea
where 
\be
{\cal D}\mu\equiv {\cal D}U \ d \lambda \ d \bar c \ d c \label{tre}
\ee
and $\delta$ denotes the nilpotent ($\delta^2=0$) BRST transformation
which acts on the links $U$ as an infinitesimal gauge transformation with
parameter $c$ and:
\bea
& & \delta \bar c = i\lambda  \nonumber \\
& & \delta c = {1\over 2} c c \label{quattro}\\
& & \delta \lambda =0 \nonumber
\eea
In eq.(\ref{due}) both the measure and the integrand are BRST invariant.
The expectation value $\langle {\cal O} \rangle$ computed through
eq.(\ref{due}) is formally equal to the one computed through eq.(\ref{uno}). 

However let us examine the quantity:
\be
F_{\cal O}(t) \equiv \int {\cal D}\mu \ e^{-S(U)}
e^{-{\alpha \over 2}\int \lambda^2} \
\exp(t \ \delta \int f \bar c) \
{\cal O}(U)\label{cinque}
\ee
which coincides with the numerator in eq.(\ref{due}), when the
real parameter $t$ is equal to $1$. We have:
\bea
& & {d F_{\cal O}(t) \over dt}=\int {\cal D}\mu \ [\delta \int f \bar c]\
e^{-S(U)} e^{-{\alpha \over 2}\int \lambda^2} \
\exp(t \ \delta \int f \bar c) \
{\cal O}(U)= \label{sei}\\
& & =\int {\cal D}\mu \ \delta \left[\int f \bar c \
e^{-S(U)} e^{-{\alpha \over 2}\int \lambda^2} \
\exp(t \ \delta \int f \bar c) \
{\cal O}(U)\right]=0 \nonumber
\eea
because the integral of a total BRST variation vanishes identically.
On the other hand we also have:
\be
F_{\cal O}(0)= \int {\cal D}\mu \ e^{-S(U)}
e^{-{\alpha \over 2}\int \lambda^2} \
{\cal O}(U)=0 \label{sette}
\ee
according to Berezin rules of grassmannian integration, because
there are no ghost or anti-ghost variables in
the integrand of eq.(\ref{sette}). Eqs.(\ref{sei}) and (\ref{sette})
imply:
\be
F_{\cal O}(1)=0 \label{otto}
\ee
Similar considerations show that:
\be
{\cal Z'}=0 \label{nove}
\ee
thus reducing eq.(\ref{due}) to an indeterminate form ${0 \over 0}$.

We stress that these manipulations are completely justified
by the finite dimensionality of the regularized functional integral.

\section{The nature of the paradox} \label{second}

In order to visualize where the problem comes from, we consider a toy
(abelian) model with only one degree of freedom, one "link" variable $U$,
which we choose to parametrize through its phase, as:
\be
U=e^{iaA} \label{link}
\ee
In eq.(\ref{dieci}), $A$ is a variable with values ranging between
$\pm {\pi \over a}$ and $a$ is a parameter, reminiscent of
the lattice spacing, in more realistic situations, whose limit
$a \rightarrow 0$ will be used to connect the periodic, compact case to the
non compact one.
The integral we consider is the trivial one:
\be
{\cal N} \equiv \int_{-{\pi \over a}}^{{\pi \over a}} dA \label{dieci}
\ee
"Gauge fixing" proceeds exactly as in section \ref{first}:
we choose a periodic function $f(A)$ and consider:
\be
{\cal N'}= \int_{-{\pi \over a}}^{{\pi \over a}} dA
\int_{-\infty}^{+\infty} d\lambda \int
d\bar c \ d c \ e^{-{\alpha \over 2} \lambda^2}
\exp(\delta [\bar c f(A)])\label{undici}
\ee
where $\delta$ is the BRST variation ($\delta^2=0$), defined by:
\bea
& & \delta A= c \label{dodici}\\
& & \delta \bar c =i\lambda \nonumber\\
& & \delta c=0 \nonumber\\
& & \delta \lambda=0\nonumber
\eea
$\delta c=0$ is of course peculiar to the present abelian
situation.

Eqs.(\ref{undici}) and (\ref{dodici}) give:
\bea
& & {\cal N'}= \int_{-{\pi \over a}}^{{\pi \over a}} dA
\int_{-\infty}^{+\infty} d\lambda \int
d\bar c \ d c \ e^{-{\alpha \over 2} \lambda^2}
e^{i\lambda f(A)} e^{-\bar c f'(A) c}=\label{tredici}\\
& & =\int_{-{\pi \over a}}^{{\pi \over a}} dA
\int_{-\infty}^{+\infty} d\lambda \
\ e^{-{\alpha \over 2} \lambda^2}
e^{i\lambda f(A)}f'(A) =\nonumber\\
& & =\sqrt{{2 \pi \over \alpha}} \int_{-{\pi \over a}}^{{\pi \over a}} df(A) \
e^{-{f(A)^2 \over 2\alpha}}=0\nonumber
\eea
for a periodic $f(A)$.

Why must we choose a periodic $f(A)$?

Because we want the expectation value of a BRST variation to vanish
identically, as a consequence of BRST invariance.

Let us choose, for instance:
\be
\Gamma \equiv \delta [\bar c F(A)]=i\lambda F(A) -
\bar c F'(A) c \label{quattordici}
\ee
A simple computation gives:
\bea
& & \langle \Gamma \rangle \equiv \int_{-{\pi \over a}}^{{\pi \over a}} dA
\int_{-\infty}^{+\infty} d\lambda \int
d\bar c \ d c \ e^{-{\alpha \over 2} \lambda^2}
e^{i\lambda f(A)} e^{-\bar c f'(A) c} \ \Gamma =\label{quindici}\\
& & = \sqrt{{2 \pi \over \alpha}} \int_{-{\pi \over a}}^{{\pi \over a}}
dA {d \over dA} (F(A) \ e^{-{f(A)^2 \over 2\alpha}}) \nonumber
\eea
Eq.(\ref{quindici}) shows that $\langle \Gamma \rangle$ is zero, only if both
$f(A)$ and $F(A)$ are periodic functions.

Choosing $\alpha=0$ in eq.(\ref{tredici}), we get:
\be
{\cal N'}=2\pi \int_{-{\pi \over a}}^{{\pi \over a}} dA
\ f'(A) \ \delta(f(A))\label{sedici}
\ee
which discloses the Gribov-like nature of the zero obtained in
eq.(\ref{tredici}): eq.(\ref{sedici}) tells us that ${\cal N'}$
gets contributions only from the zeros of $f(A)$. Each zero contributes
alternatively a factor $\pm 1$ and the periodicity of $f(A)$ implies an
even number of zeros.

\section{The solution} \label{third}

There is, however, a way out which exploits the compact nature of the
variable $A$. The basic idea is simple: it amounts to change the
BRST formulation, so that in eq.(\ref{sedici}) the periodic
$\delta$-function appears. As we will show in a moment, this allows the
choice of a function $f(A)$ periodic up to a shift, i.e.:
\be
f(A+{2\pi \over a}) = f(A)+{2\pi k\over a}\label{diciassette}
\ee
with an integer $k$, still leaving a periodic integrand in eq.(\ref{sedici}).
In this way the number of zeroes is odd and the Gribov problem is overcome.

To be more precise, let us start recalling some properties of the periodic
$\delta$-function, $\delta_P(x)$. $\delta_P(x)$ is defined as\cite{light}:
\be
\delta_P(x)\equiv \sum_{n=-\infty}^{\infty}
\delta(x-n{2\pi \over a}) \label{diciotto}
\ee
and may also be represented as:

\be
\delta_P(x)= {a \over 2\pi}\sum_{n=-\infty}^{\infty}
e^{i n a x} \label{diciannove}
\ee

The modification I propose, is to allow the variable $\lambda$ in
eq.(\ref{undici}) to assume discrete values only:
\be
\lambda_n\equiv a n \label{venti}
\ee
with integer $n$, and replace eq.(\ref{undici}) by:
\be
{\cal N''}= a \sum_{n=-\infty}^{+\infty}
\int_{-{\pi \over a}}^{{\pi \over a}} dA
\int d\bar c \ d c \ e^{-{\alpha \over 2} \lambda_n{}^2}
\exp(\delta [\bar c f(A)])\label{ventuno}
\ee
The BRST transformations have the same form as those defined in
eq.(\ref{dodici}), only with $\lambda$ replaced by $\lambda_n$.
The same steps leading to eq.(\ref{tredici}) give now:
\be
{\cal N''}= a \sum_{n=-\infty}^{+\infty}
\int_{-{\pi \over a}}^{{\pi \over a}} dA
\ e^{-{\alpha \over 2} \lambda_n{}^2}
e^{i\lambda_n f(A)}f'(A) \label{ventidue}
\ee
We can now choose a function $f(A)$ obeying the condition stated in
eq.(\ref{diciassette}): while $f(A)$ itself cannot be expressed
globally in terms of the link $U$, defined in eq.(\ref{link}),
both $e^{i\lambda_n f(A)}$ and $f'(A)$ can, being strictly periodic functions
of $A$. We can easily check that the consequences expected
from BRST invariance are satisfied by the new definition,
eq.(\ref{ventidue}). In particular the expectation value of $\Gamma$,
defined in eq.(\ref{quattordici}), still vanishes for a periodic $F(A)$.
However, in general, ${\cal N''}\neq 0$:
the argument of ref.\cite{neub1}, reviewed in section \ref{first},
fails to apply in this case because, while $e^{i\lambda_n f(A)}$ is a
periodic function, $e^{it\lambda_n f(A)}$ is not periodic, except for integer
values of $t$.

It is worth noticing that, in the "continuum limit",
$a \rightarrow 0$, eq.(\ref{ventidue}) converges to the usual
BRST expression:
\be
\lim_{a \rightarrow 0} {\cal N''} =
\int_{-\infty}^{+\infty} dA \int_{-\infty}^{+\infty}
d \lambda\ e^{-{\alpha \over 2} \lambda^2}
e^{i\lambda f(A)}f'(A) \label{ventitre})
\ee
thus recovering a continuous valued $\lambda$ variable.

The above arguments seem to translate directly into
the much more complicated case of lattice gauge theories where we
have a set of Lagrangian multipliers, $\lambda(x)$, for each space-time point
$x$. The basic move
is the discretization of the range of the Lagrange multipliers, $\lambda_n(x)$,
which allows the introduction of "twisting" gauge fixings, as exemplified in
eq.(\ref{diciassette}). We stress again that this amounts to a non
conventional choice of the gauge condition, $f(A,x)$, which should not be
expressible in terms of the link variables, $U(x)=e^{iaA(x)}$, while
$e^{i\int d^4x \lambda_n(x)f(A,x)}$ should.

In realistic cases, of course, it is not
easy to identify gauge-fixing conditions apt to avoid the vanishing results
implied by eqs.(\ref{otto}) and (\ref{nove}): it is, however, most
encouraging that, in presence of discretized $\lambda_n$'s,
eqs.(\ref{otto}) and (\ref{nove}) cannot be shown to be valid any more.

\section{Conclusions} \label{fourth}

We have proposed a possible way out to an old
paradox concerning non-perturbative BRST symmetry
in lattice regularized gauge theories.
The solution consists in changing the BRST formulation
by allowing the Lagrange multipliers, which enforce the gauge condition, to
range over discrete values only. This modification gives more freedom
in the choice of the gauge condition, allowing, in simple examples, to
circumvent the paradox\footnote{The possibility of a multivalued
gauge-fixing condition is hinted in the concluding
remarks of ref.\cite{neub1}, but it is not pursued any further.}.
Even more convincing is the fact that, within this modified
BRST formulation, the steps leading to the paradox cannot be implemented.

In the continuum limit, the modified BRST formulation naturally converges
towards the conventional continuum one, in the sense that the discretized
Lagrange multipliers become continuous variables as $a \rightarrow 0$.

It is hard to think that the modified BRST will have any impact on actual
numerical simulations, because its implementation seems rather involved.
To this we must add that Neuberger's argument, although relevant to
gauge-fixings used in practice, was always ignored by the practitioners
in the field.

Are there any relevant consequences for the Rome approach?
I don't think so. In the Rome approach one starts from a
discretized theory which is not gauge invariant and tries to merge it,
as $a \rightarrow 0$, into a {\it continuum} gauge fixed, gauge invariant
(i.e. satisfying the {\it continuum} BRST identities) theory. As stressed
several times, within the Rome approach one does not deal with regularized,
exactly gauge-invariant systems.

\section*{Acknowledgements}

I thank the CERN Theory Division for the kind hospitality.

\noindent I also thank G. C. Rossi for discussions.

\end{document}